\documentclass[10pt]{iopart}
\usepackage{graphicx}
\usepackage{bm}
\usepackage[T1]{fontenc}
\usepackage{mathptmx}
\usepackage{etoolbox}
\usepackage{cite}

\usepackage{multirow}
\usepackage{hyperref} 
\usepackage{makecell} 
\usepackage{amssymb} 
\usepackage{amsmath,mathtools}
\usepackage{wasysym} 
\usepackage{gensymb} 
\usepackage{enumerate}

\usepackage{upgreek} 

\begin{document}

\title[High resolution spatially extended 1D laser scattering diagnostics using volume Bragg grating notch filters]{High resolution spatially extended 1D laser scattering diagnostics using volume Bragg grating notch filters}

\author{Junhwi Bak$^{1*}$, Jean Luis Suazo Betancourt$^2$, Anuj Rekhy$^1$, Amirhossein Abbasszadehrad$^1$, Richard B. Miles$^1$, Christopher M. Limbach$^1$, and Mitchell L. R. Walker$^2$}
\address{$^1$ Department of Aerospace Engineering, Texas A$\&$M University, College Station, Texas $77843$, USA}
\address{$^2$ Daniel Guggenheim School of Aerospace Engineering, Georgia Institute of Technology, North Ave NW, Atlanta, GA $30332$, USA}
\ead{junhwib@tamu.edu}

\vspace{10pt}
\begin{indented}
\item[]{\today}
\end{indented}

\begin{abstract}
Laser light scattering systems with volume Bragg grating (VBG) filters, which act as a spectral/angular filter, have often been used as a point measurement technique with spatial resolution as low as a few hundred $\upmu$m  defined by the beam waist. In this work, we demonstrate how VBG filters can be leveraged for spatially resolved measurements with several $\upmu$m resolution perpendicular to the laser propagation axis over a few mm along the beam propagation axis. The rejection ring, as determined by the angular acceptance criteria of the filter, is derived analytically, and the use of the ring for 1D laser line rejection is explained. For the example cases presented, having a focused probe beam waist with a diameter of $\sim 150$~$\upmu$m, the rejection ring can provide up to several mm length along the beam propagation axis for a 1D measurement, which is also tunable. Additionally, methods to further extend the measurable region are proposed and demonstrated; using a collimation lens with a different focal length or using multiple VBG filters. The latter case can minimize the scattering signal loss, without the trade-off of the solid angle. Such use of multiple VBGs is to extend the measurable region along the beam axis, which differs from the commonly known application of multiple filters to improve the suppression of elastic interferences. 1D rotational Raman and Thomson scattering measurements are carried out on pulsed and DC discharges to verify this method. The system features compactness, simple implementation, high throughput, and flexibility to accommodate various experimental conditions.

\end{abstract}

\vspace{2pc}

%

\ioptwocol
\section{\label{sec:intro}Introduction}

Laser light scattering is an important diagnostic tool. It allows non-perturbative access to physically harsh test articles in high-pressure/high-temperature environments, where measurements by a physical probe are practically limited. Accurate measurement of the relevant state quantities of the scattering gas, such as temperature and density, allow for the validation of physical simulation models, ultimately leading to a better understanding of physics and better design of specific applications. This is due to the minimal assumptions required in order to employ these laser light scattering techniques, which may not even require local thermal equilibrium.

Scattering configurations are typically off-axis of the laser beam propagation axis, with the practical maximum signal being collected perpendicular to the beam polarization direction. The laser-focused beam waist is strongly localized, providing high spatial resolution that is set by the beam waist and another limiting aperture. This configuration produces a localized measurement in contrast with path-integrated techniques such as optical emission spectroscopy and microwave interferometry which require tomographic methods to spatially resolve the measurements. Along the laser beam, the spatial resolution is determined by the optics and a detection camera's resolution in the case of a free space optical detection system. This can be several tens of $\upmu$m. Meanwhile, the maximum resolution perpendicular to the beam is determined by the beam waist, which is typically about 100 - 200 $\upmu$m.

Different scatterers in the detection volume, such as monatomic and polyatomic neutral gas particles, ions and electrons, will scatter differently. Knowledge of the scatterers and detection parameters allow for direct measurement of some of their key state properties. Rayleigh scattering is elastic light scattering by the induced dipole moment from bound electrons and can be used to measure the gas temperature, velocity, and density.\cite{Miles2001LaserScattering} Raman scattering, an inelastic scattering process associated with the transition in a molecule's rotational or vibrational state, provides molecular number densities and the rotational/vibrational temperatures. Thomson scattering from free electrons allows measuring electron temperature and density and the electron energy distribution function\cite{Milder2019ImpactScattering}, fundamental properties that define the composition of plasma\cite{Sheffield2011PlasmaRadiation}.

The intensity of the elastically-scattered light (Rayleigh scattering and stray light) is several orders higher than that of Raman or Thomson scattering. Differential scattering cross-sections $d\sigma/d\Omega$ of different particles for the right-angle-scattering\cite{Hubner2017ThomsonJets} in $10^{-32}$ m$^2$sr$^{-1}$ are 3.9 for Rayleigh (N$_2$), 0.054 for Raman (N$_2$ $J$-$J'$=$6$-$8$), and 794 Thomson (e$^-$).  It is also noteworthy that the intensity of the stray light can also be significant close to the laser line because of reflections and other factors that allow this light to make it through the detection system if left unaddressed. The Thomson signal in many plasma applications such as low-temperature weakly-ionized plasma becomes difficult to detect because the neutral density is several orders of magnitude greater than the electron density. Thus, to detect Raman and Thomson scattering signals, the light near the laser line (Rayleigh scattering/stray light) should be effectively removed to prevent saturation of the detector. Such laser line rejection can be performed with multiple techniques such as a triple grating spectrograph\cite{Sande2002LaserRejection,Tomita2014MeasurementsScattering,Carbone2015ThomsonChallenges,Sasaki2015ElectronMPa,Hubner2017ThomsonJets,Roettgen2016Time-resolvedHelium,Obrusnik2016CoherentBehaviour,Tomita2020MeasurementAir,Tan2021ElectronDiagnosis}, a vapor cell\cite{Bakker2001ThomsonColumn, Lee2002SpectrallyFilter, Limbach2014RayleighPlasmas}, a glass/interference filter\cite{Weeling1996}, a physical mask\cite{Chen2019Time-resolvedDischarge,Wu2021InvestigationScattering}, and a volume Bragg grating (VBG) filter \cite{Paillet2010HighFilters,Glebov2012,Klarenaar2015Note:Filter,Vincent2018AStudies,Wu2020,Wu2021,Slikboer2021ImpactJet,Yatom2022CharacterizationBroadening}. 

Past data analysis from laser scattering experiments using the VBG filter were effectively limited to point measurements \cite{Paillet2010HighFilters, Glebov2012, Klarenaar2015Note:Filter, Vincent2018AStudies,Yatom2022CharacterizationBroadening,Klarenaar2018HowScattering}, which may have been a significant limitation of the method. Recent works by Wu \etal\cite{Wu2021} present one-dimensional (1D) measurements of Thomson scattering of a nanosecond repetitively-pulsed discharge using a VBG filter, showing the possible extension of spatial measurement of light scattering. However, no detailed explanation of the use is given.

In this paper, we elaborate on the details of how to realize 1D measurement with VBG filters based on the first principles of VBG filter rejection characteristics. Several useful equations to characterize the rejection region are provided. The equations help determine design parameters such as a measurable length and a resolution for the design stage of such a system. Additionally, we propose a method to further extend the measurement region using multiple VBGs. It should be noted that this proposed method differs from the commonly known application of multiple VBGs to improve the rejection capability. The extended measurement volume along the laser propagation axis by VBG filters is expected to be greatly advantageous by maintaining the desired spatial resolution perpendicular to the beam propagation axis for applications that require high throughput light scattering diagnostics.

\section{\label{sec:principle}Principle of volume Bragg grating filters}
\subsection{\label{sec:bg}General background}
The VBG is a diffractive grating that rejects light at certain wavelengths and given angular conditions by refractive index modulation in the volume of a photosensitive material. Depending on the design of the diffraction angle/orientation/modulation, VBGs can work in several types, such as a transmitting Bragg grating (TBG), a reflecting Bragg grating (RBG), or a chirped Bragg grating (CBG). Among these VBGs, RBGs exhibit two characteristic features, a narrow band spectral filter and an angular filter. Configured as a narrow-band notch filter, RBGs have spectral bandwidths as narrow as $5$ cm$^{-1}$.\cite{Glebov2012} A filter allows the selective reflection of light at a specifically designed wavelength incidence at the Bragg condition, thus, performing a dual function as an angular and a spectral filter. A detailed theoretical model of volume gratings can be found elsewhere.\cite{Ciapurin2005ModelingGlass,Ciapurin2005ModelingGlassb} 

The use of the reflective VBG as a notch filter for light scattering diagnostics provides several advantages. First, the optical systems required for signal collection are significantly simplified ; in the simplest case, the optical system consists of two lenses, a VBG filter, a single grating spectrometer, and an intensified detector. This naturally leads to a significant improvement in the system's total efficiency through the use of minimal optics. Additional throughput is gained through the transmittance of the filter outside of the spectral rejection region, with transmittance of the unfiltered wavelengths being up to 95\%. Additionally, the method provides excellent flexibility in the system design. Depending on the necessity of a specific experiment, simply adding an additional filter can improve the total laser line rejection and spatial extension of the measurement region while maintaining throughputs that are an order of magnitude higher than that of multi-stage spectrometers.

\begin{figure}[b]
	\centering
	\includegraphics[width=0.7\linewidth]{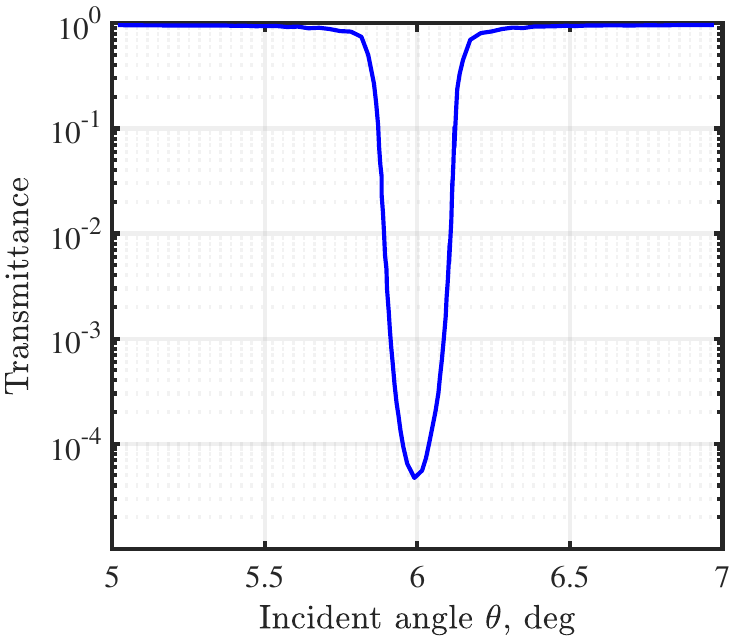}
	\caption{\label{fig:BNF} Typical angular spectrum for an OD4 reflecting volume Bragg garting filter (Data from OptiGrate).}
\end{figure}

Commercially available reflecting volume Bragg gratings used as notch filters show spectral bandwidths as narrow as $5$ cm$^{-1}$.\cite{Glebov2012} A typical transmittance curve of an OD4 filter designed for 532 nm (BNF-532; OptiGrate) as a function of incident angle to the filter surface is shown in Fig.~\ref{fig:BNF}. The curve data are read from a specification sheet provided by OptiGrate. Approximately within $0.1\degree$ of the designed input angle $6\degree$, the filter provides $10^{-4}$ attenuation. This angular spectra characteristic is a key element to realize a 1D measurement with BNFs for light scattering diagnostics. 

\subsection{\label{sec:ring}Evaluation of a rejection ring and manipulation of the rejection ring position}
The transmittance curve as a function of incident angle to the filter surface, in Fig.~\ref{fig:BNF} shows that $10^{-4}$ transmission is attainable $6\pm 0.05\degree$ and $10^{-3}$ transmission is attainable $6\pm 0.09\degree$. This indicates that only a certain \textit{spatial range} of light from the object plane can be rejected by the filter if the incident angle of the approximately collimated light falls within the ranges given above. We define the rejecting angle $\theta_\mathrm{r}$, and OD4 and OD3 range as $\delta\theta_\mathrm{OD4}$ and $\delta\theta_\mathrm{OD3}$, respectively. $\theta_\mathrm{r}=6\degree$, $\delta\theta_\mathrm{OD4}=0.05\degree$, and $\delta\theta_\mathrm{OD3}=0.09\degree$ are used as example values for a filter in the following evaluation.

\begin{figure}
	\includegraphics[width=0.99\linewidth]{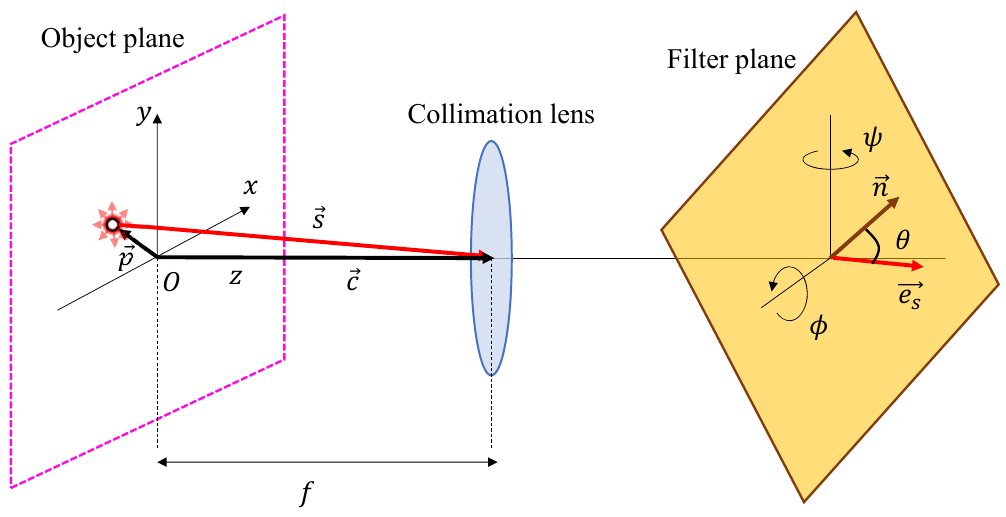}
	\caption{\label{fig:schema} Schematics of an optical arrangement between an object plane, a collimation lens, and a filter plane.}
\end{figure}

Let us consider an optical arrangement of a collimating lens and a VBG as shown in Fig.~\ref{fig:schema}, where
\begin{itemize}
	\item $\vec{c}$; a position vector of the center of the collimation lens, $\vec{c} = \left[0,0,f\right]^\intercal$ in the $x$-$y$-$z$ coordinate, where $f$ is a focal length of the collimation lens.
	\item $\vec{p}$; a position vector of a point $p$ on the object plane.
	\item $\vec{s}$; a vector from the point $p$ to the center of a collimation lens, $\vec{s} \equiv \vec{c} - \vec{p}$.
	\item $\vec{e}_s$; a unit vector of $\vec{s}$, $\vec{e}_s \equiv \vec{s}/|\vec{s}|$.
	\item $\vec{n}$; a unit normal vector of the Bragg notch filter plane. The initial orientation $\vec{n}_\mathrm{i}=\left[0,0,1\right]^\intercal$.
\end{itemize}

For a scattering source $p$ on the object plane as a point source, the light scattered by the source is collimated by the lens, assuming that the lens is one focal length away from the object plane. Note that the direction of the collimated rays is determined by a ray directly pointing at the center of the collimation lens from the source $p$. This collimated ray vector is denoted as $\vec{s}$. Then, the angle $\theta$ between $\vec{s}$ and $\vec{n}$ becomes the incident angle of the ray to the filter, which can be obtained by,
\begin{equation}
\theta = \cos^{-1} \left( {\vec{e}_s \cdot \vec{n}} \right). \label{eq:theta}
\end{equation}
Thus, by calculating $\theta$ for each location on the object plane and taking into account the transmittance from Fig.~\ref{fig:BNF}, spatial distribution of a rejected location can be identified, which eventually turns out to be a ring shape. We term this rejected area as the \textit{rejection ring}, the justification for which will be shown below.  It is noteworthy that the described spatial rejection region analysis can be fundamentally applicable to any angular/spectral filters such as interference filters or holographic spectral filters. Using the angular spectra given in Fig.~\ref{fig:BNF}, a rejection ring (center) diameter $D_\mathrm{r}$ can be estimated as,
\begin{equation}
D_\mathrm{r} = 2f\tan{\theta_\mathrm{r}}, \label{eq:D_r}
\end{equation}
and the thickness of the OD4 ring  $t_\mathrm{r,OD4}$ is,
\begin{equation}
t_\mathrm{r,OD4} = f\left( \tan{ \left( \theta_\mathrm{r}+\delta\theta_\mathrm{OD4}\right)} -  \tan{\left( \theta_\mathrm{r}-\delta\theta_\mathrm{OD4}\right)}\right), \label{eq:t_OD4}
\end{equation}
and the thickness of the OD3 ring  $t_\mathrm{r,OD3}$ is,
\begin{equation}
t_\mathrm{r,OD3} = f\left( \tan{ \left( \theta_\mathrm{r}+\delta\theta_\mathrm{OD3}\right)} -  \tan{\left( \theta_\mathrm{r}-\delta\theta_\mathrm{OD3}\right)}\right). \label{eq:t_OD3}
\end{equation}
Note that different VBGs, having different angular rejection properties, will have different rejection incident angles and OD4/OD3 ranges that can be obtained from their specifications. The reason for evaluating the two OD rings is because one can choose either ring for a diagnostic system design depending on the attenuation requirement of the test condition in a trade-off between the attenuation and the spatial rejection size, as well as other factors like cost and lead time. 

\begin{figure}
	\includegraphics[width=0.99\linewidth]{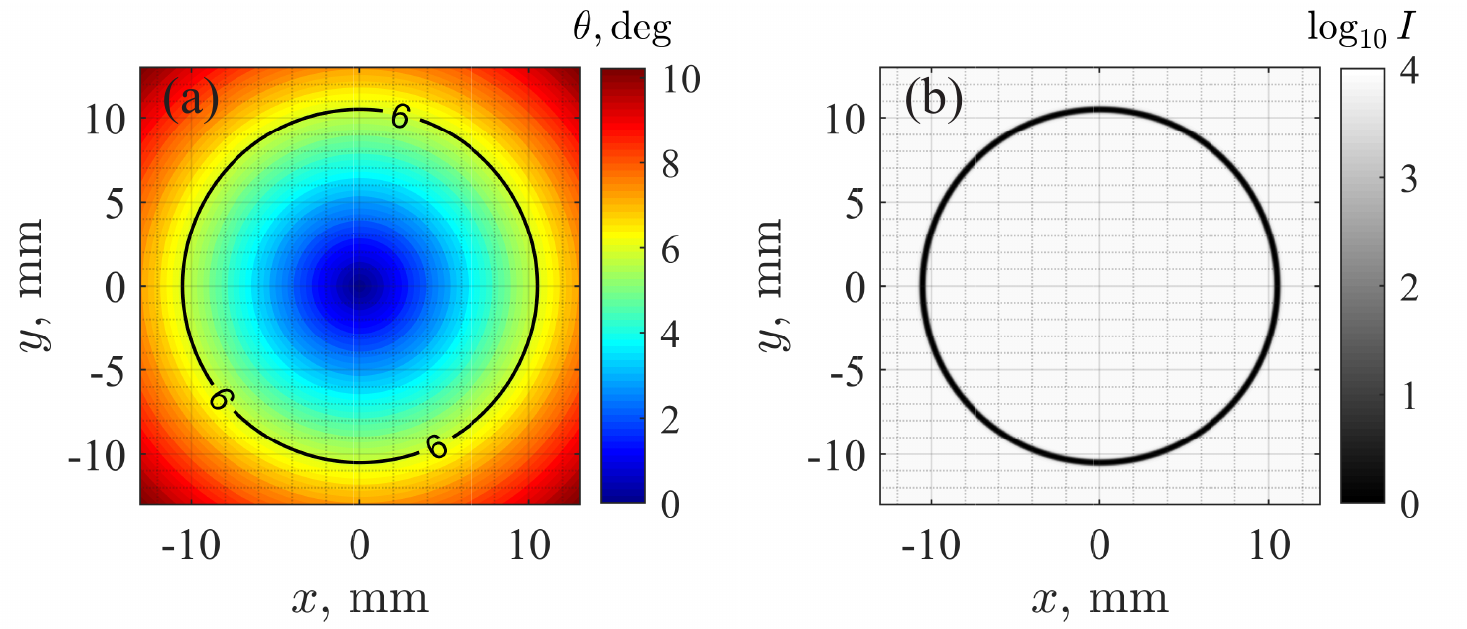}
	\caption{(a) a map of incident angle $\theta$ to the filter surface depending on the position of the scattering source on the object plane. (b) a map of light intensity $I$ where the initial intensity $I=10^4$ was over the whole object plane. The attenuation by the filter on a specific ring-shape region is observable.}
	\label{fig:map_sam}
\end{figure}

Figure~\ref{fig:map_sam} shows a sample calculated incident angle distribution (a) and a light intensity map (b). For this calculation, $f=100$ mm is used, and an initial intensity, $I=10^4$, is given for every location on the object plane. The position having the incident angle $\theta=6\degree$ is indicated as a black solid line in (a), and accordingly, the rejected region appears in a shape of a ring in (b). For the intensity map, the angular transmittance curve in Fig.~\ref{fig:BNF} is used. $D_\mathrm{r}\approx 21$ mm, $t_\mathrm{r,OD4}\approx 177$ $\upmu$m, and $t_\mathrm{r,OD3}\approx 318$ $\upmu$m is obtained.

Now we consider angle tuning of the filter plane with two degrees of freedoms,a rotation with respect to the $x$ axis (pitch control) and a rotation with respect to the $y$ axis (yaw control) using Euler angles. Note that the rotation along $z$ axis has no effect due to the axisymmetric relation, which is enforced through colinearity of the optics (placing the optics along the same optical axis). We can determine the angle of any vector from one plane (or frame of reference) to any other plane by relating them through a matrix transformation using rotation matrices. 

Defining a rotation matrix for each of the angles above, one with respect $x$ axis (pitch) with an angle $\phi$ as,
\begin{equation}
R_x(\phi) \equiv \begin{bmatrix}
1 &           0 &             0\\
0 &  \cos{\phi} &   -\sin{\phi}\\
0 &  \sin{\phi} & ~~~\cos{\phi} 
\end{bmatrix},
\end{equation}
and a rotation matrix of the angle with respect to $y$ axis with an yaw angle $\psi$ as ,
\begin{equation}
R_y(\psi) \equiv \begin{bmatrix}
~~~\cos{\psi} & 0 & -\sin{\psi}\\
0 & 1 &            \\
-\sin{\psi} & 0 &  ~~~\cos{\psi} 
\end{bmatrix},
\end{equation}
the normal vector of the filter $\vec{n}$ is obtained as,
\begin{equation}
\vec{n} = R_x(\phi) R_y(\psi) \vec{n}_\mathrm{i}
\end{equation}
Then, a final angle $\theta$ that rays from the point $p$ form with respect to the VBG can be obtained by Eq.~(\ref{eq:theta}).

\begin{figure}
	\includegraphics[width=0.99\linewidth]{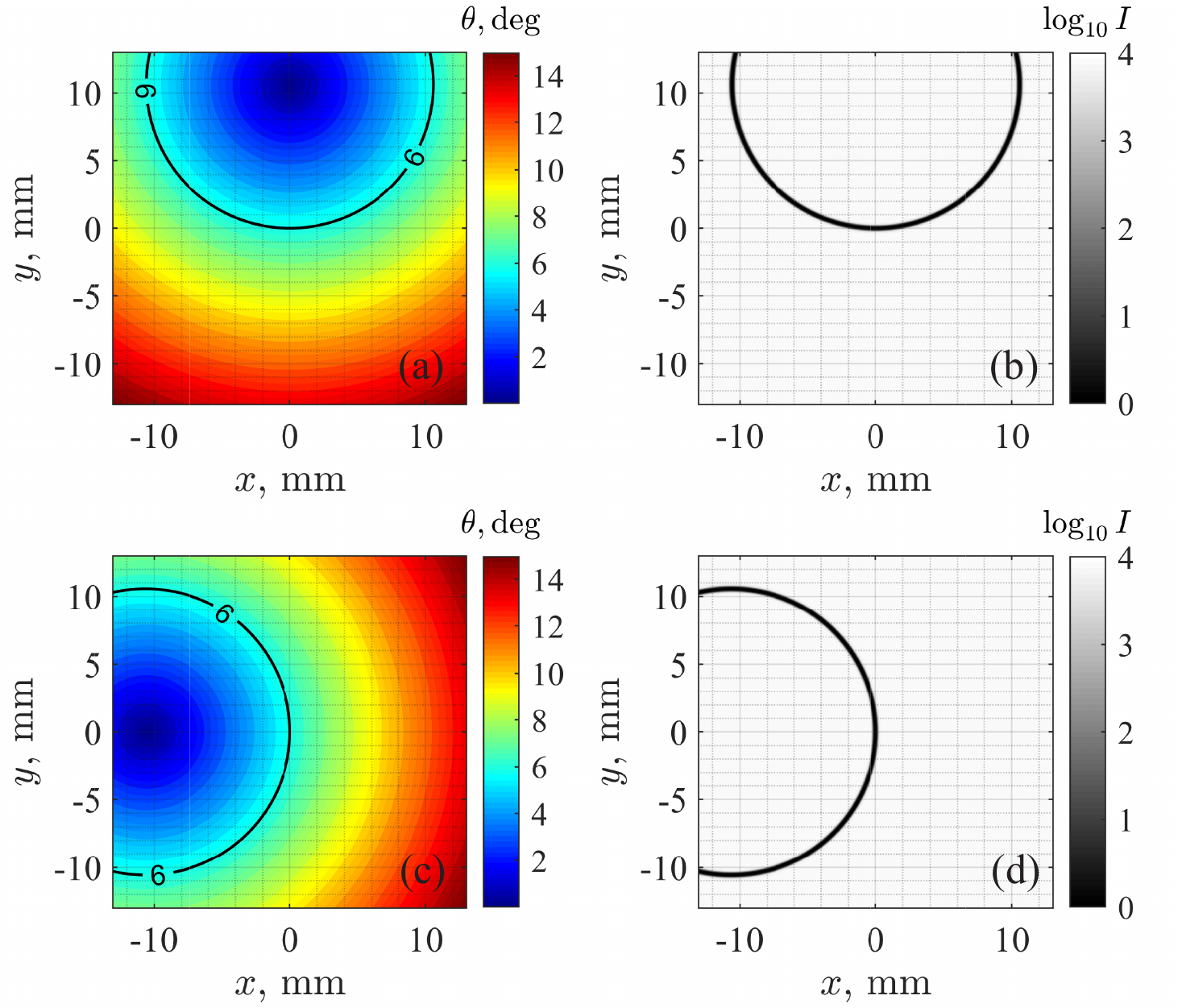}
	\caption{(a) a map of incident angle $\theta$ for a pitch angle $\phi=6\degree$ (b) a map of light intensity $I$ for $\phi=6\degree$. (c) $\theta$ for a yaw angle $\psi=6\degree$ (d) $I$ for $\psi=6\degree$.}
	\label{fig:yaw_pitch} 
\end{figure}

Influence of the pitch/yaw angle adjustment are shown in Fig.~\ref{fig:yaw_pitch}, with (a) being a map of incident angle $\theta$ for a pitch angle $\phi=6\degree$, (b) a map of light intensity $I$ for $\phi=6\degree$, (c) a map of incident angle $\theta$ for a yaw angle $\psi=6\degree$, and (d) a map of light intensity $I$ for $\psi=6\degree$. Note that a $\phi$ control moves the rejection ring up and down along the y axis, and a $\psi$ control moves the image's rejection ring left and right along the x axis. 

For rejection, we are concerned with the relative size and location of the rejection ring on the image of the laser beam region, imaged at our detector plane. This is because without any rejection, the detector saturates due to redistribution of the strong laser line light from Rayleigh scattering as well as reflections through the collection system. Therefore, the laser line light must be adequately blocked in the field of view of our detector in order to be able to collect the relatively weak rotational Raman and Thomson scattering light. For a given VBG, collimating/objective lens and focusing/imaging lens, the size of the rejection ring on the image of the laser beam at the imaging plane as well as its position on the laser beam image can be directly controlled by controlling the pitch and yaw angles, assuming good collimation through the collimating lens. This can be further controlled with a second VBG.

\section{\label{sec:1Dapp}Application of the rejection ring to 1D measurement}
\subsection{\label{sec:align}Alignment of a VBG filter with respect to a laser line}
\begin{figure}
	\includegraphics[width=0.99\linewidth]{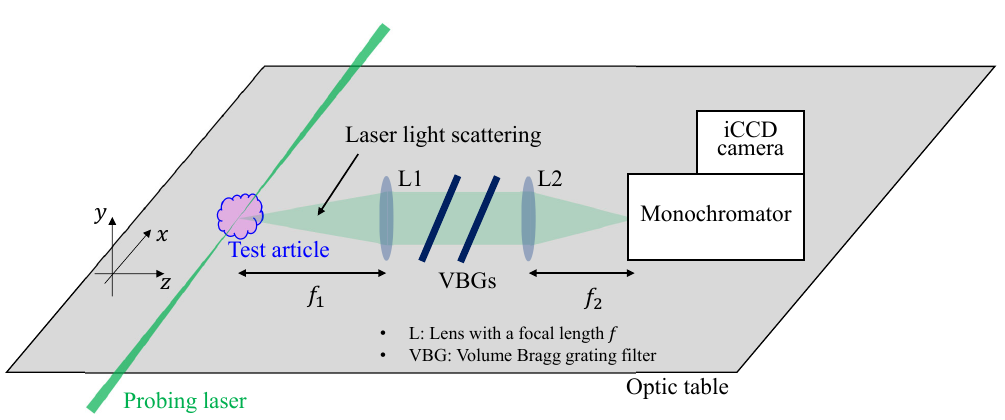}
	\caption{Schematic of a typical optical setup for laser light scattering diagnostics with a volume Bragg grating filter.}
	\label{fig:optics}
\end{figure}

A typical optical setup for laser light scattering diagnostics with a volume Bragg grating is shown in Fig.~\ref{fig:optics}; a probing laser propagating along the $x$-axis (with vertical polarization in $y$) on the $x$-$z$ plane, and light scattering collection in $z$ direction (90\degree~ collection).

\begin{figure}
	\includegraphics[width=0.99\linewidth]{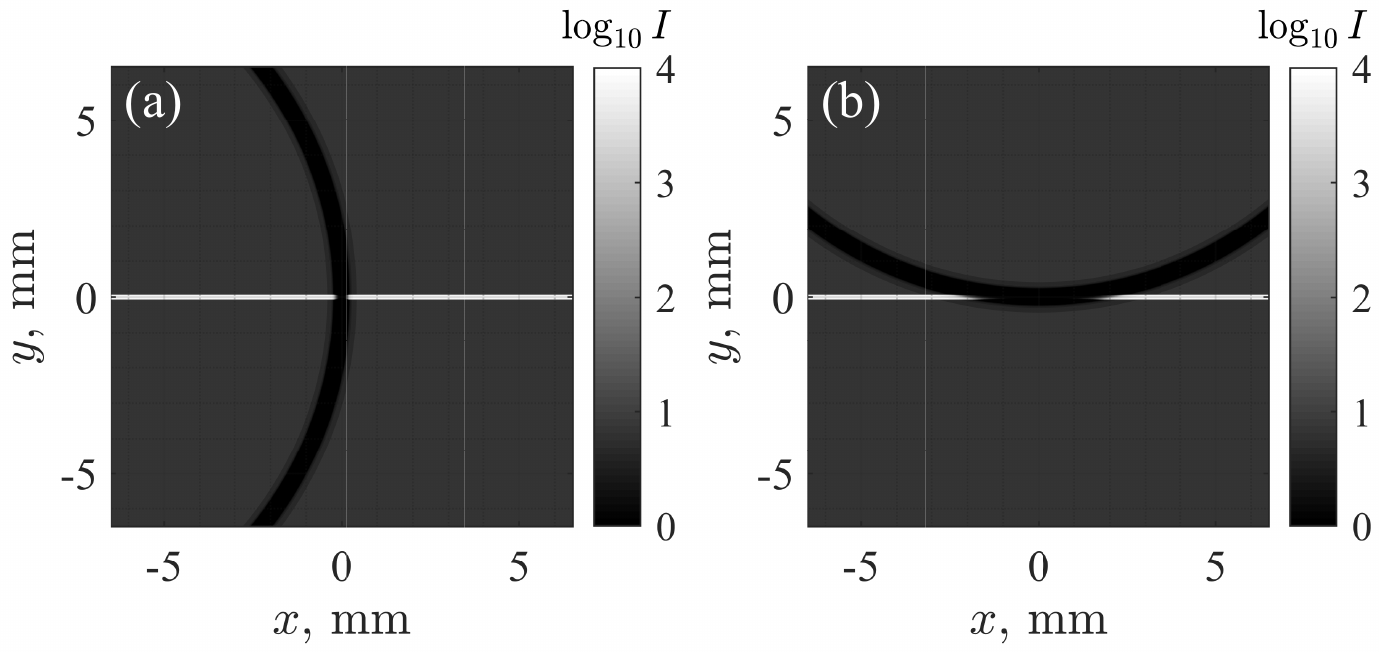}
	\caption{Laser line rejection by tuning the VBG angle (a) $\left( \phi,\psi \right)=\left(0\degree,6\degree\right)$ and (b) $\left( \phi,\psi \right)=\left(6\degree,0\degree\right)$. The white line indicates a laser line image.}
	\label{fig:0D1D}
\end{figure}

To reject the strong laser line in scattering diagnostics, a VBG is rotated. Figure~\ref{fig:0D1D} shows simulated laser line images depending on a specific angle tuning. A laser line in $x$ direction (the white region) is simulated as a Gaussian beam focused into a beam waist of $150$ $\upmu$m and is given the intensity $I=10^4$. Note that, within the viewed region, the variation of the probe beam diameter is negligible. The intensity $I=8$ is given in the area other than the laser image, simply to provide a little contrast on the resulting intensity map. 

First, the yaw $\psi$ tuning, Fig.~\ref{fig:0D1D}(a), moves the rejection ring left or right (along $x$ axis), having the rejection ring stands perpendicular to the laser. This only rejects a small portion of the laser line (given as $t_\mathrm{r}$, in the shown example, $\sim 200$ $\upmu$m), limiting the setup to a point measurement with a resolution of $t_\mathrm{r}$. On the other hand, in the totally same optical setup, by tuning the pitch $\phi$, Fig.~\ref{fig:0D1D}(b), the rejection ring moves up or down (along $y$ axis), being tangential to the laser line. As a result, a wider spatial region (in the shown example, $\sim 2.5$ mm) falls into the rejection ring (the high OD region), allowing a 1D spatially-resolved measurement perpendicular to the laser beam propagation axis, in this case, along the y axis.

It is clear now how specific angle control can result in either a point measurement, averaged over the number of pixels blocked along the axis perpendicular to the laser beam propagation axis, or a 1D measurement, by resolving at each pixel row along this direction. Note that if a system adopts only a single angle tuning for a VBG, i.e. the yaw angle control with a rotation stage (rotation along $y$ axis) while the probing laser is propagating along an optic table, as seen in Fig.~\ref{fig:optics}, only a narrow portion of the laser is rejected, seen in Fig.~\ref{fig:0D1D} (a). This makes a laser scattering system (one where no fiber is used) to be fundamentally limited to the point measurement. It is noteworthy that the diagnostic can be easily extended from a point measurement to a 1D measurement with the exact same experimental setup. Note that the length of $\sim 2.5$~mm in the example, seen in Fig.~\ref{fig:0D1D} (b), is long enough to resolve a full radial distribution of test article properties such as thin discharge columns, microplasmas, streamers, etc. More importantly, the spatial resolution perpendicular to the beam propagation axis determined directly by the detector resolution and imaging optics, that can be as small as a few $\upmu$m and can also be easily adjusted by a choice of imaging optics. This is because each pixel row corresponds to a different spatial location on the object plane. Such resolution enables to resolve steep gradients.

\subsection{\label{sec:extend}Extended 1D measurement with a VBG}
\begin{figure}
	\includegraphics[width=0.99\linewidth]{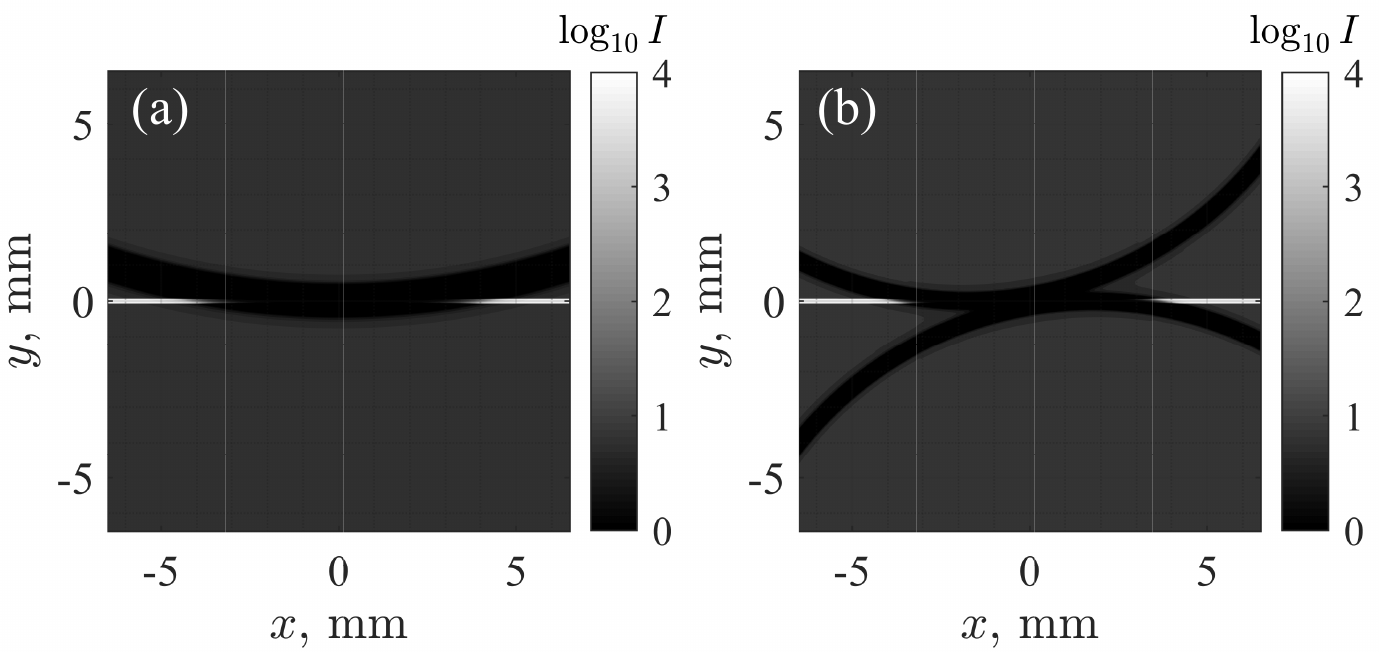}
	\caption{Laser line rejection (a) with a collimation lens focal length $f = 100$ mm with $\left( \phi,\psi \right)=\left(6\degree,0\degree\right)$ and (b) with two VBGs, where one is angle tuned at $\left( \phi,\psi \right)=\left(6\degree,1\degree\right)$ another with $\left( \phi,\psi \right)=\left(-6\degree,-1\degree\right)$.}
	\label{fig:exten}
\end{figure}

Considering the rejection ring size $D_\mathrm{r}$, or the thickness $t_\mathrm{r}$ being proportional to the collimation lens focal length $\propto f$, Eq.~(\ref{eq:D_r}-\ref{eq:t_OD3}), a quick method to extend 1D measurement region is to implement a collimation lens with a longer focal length. Figure~\ref{fig:exten}(a) shows such application with a different focal length ($f=200$~mm) of the collimation lens where a VBG is at $\left( \phi,\psi \right)=\left(6\degree,0\degree\right)$. Compared to the case of $f=100$~mm, seen in Fig.~\ref{fig:0D1D}(b), the case with a collimation lens with $f = 200$~mm, Fig.~\ref{fig:exten}(a), shows that the rejection ring becomes larger (thicker), a laser line in further wider 1D region can be rejected, extending spatial measurement domain. However, it should be noted that, as the lens is positioned further away from the scattering volume, the solid angle for the signal collection decreases; the shown examples with $f=100$~mm and $f=200$~mm, $\sim4$~times weaker signal. Thus, there exists a trade-off between measurable 1D region and total signal intensity (assuming the filter size is kept the same).

To overcome this trade-off, one can use multiple filters by simply adding an additional filter after one, see Fig.~\ref{fig:optics}. Figure~\ref{fig:exten}(b) shows laser line rejection by two VBGs; one angle tuned with $\left( \phi,\psi \right)=\left(6\degree,1\degree\right)$ another with $\left( \phi,\psi \right)=\left(-6\degree,-1\degree\right)$. Thus, the approximately two times extended 1D region $\sim 5.5$~mm, can be rejected by slightly shifting the rejection rings from each filter with $\psi$ tuning. Note that, as the total transmission of other wavelengths unaffected by the filter is typically $>90\%$, the signal loss can be significantly reduced compared to the case of using a longer focal length lens. It should be noted that this proposed method of using multiple VBGs differs from the commonly known application of multiple VBGs to improve the rejection power.

\subsection{\label{sec:ringdemo}Demonstration of laser line rejection with a VBG notch filter}
An experimental setup is similar to that shown in Fig.~\ref{fig:optics}. A Nd:YAG frequency-doubled 532 nm laser (10 Hz, 5 ns;NL315-10SH, Ekspla) is used as a laser source, and images are captured by an emiCCD camera (PI-MAX4:1024EMB, Princeton Instruments). The imaging is done with 1:1 imaging optics, and the image size is 13 mm $\times$ 13 mm for $1024\times1024$ pixels (the horizontal view is limited by a slit on the spectrometer). Note that the spectrometer (IsoPlane 320, Princeton Instruments) and camera assembly are rotated to align the spectrometer slit to be parallel with the laser, so the demonstration images to be shown are 90 deg rotated. An OD4 VBG filter designed for 532 nm (BNF-532; OptiGrate) is used.  

\begin{figure}
	\includegraphics[width=0.99\linewidth]{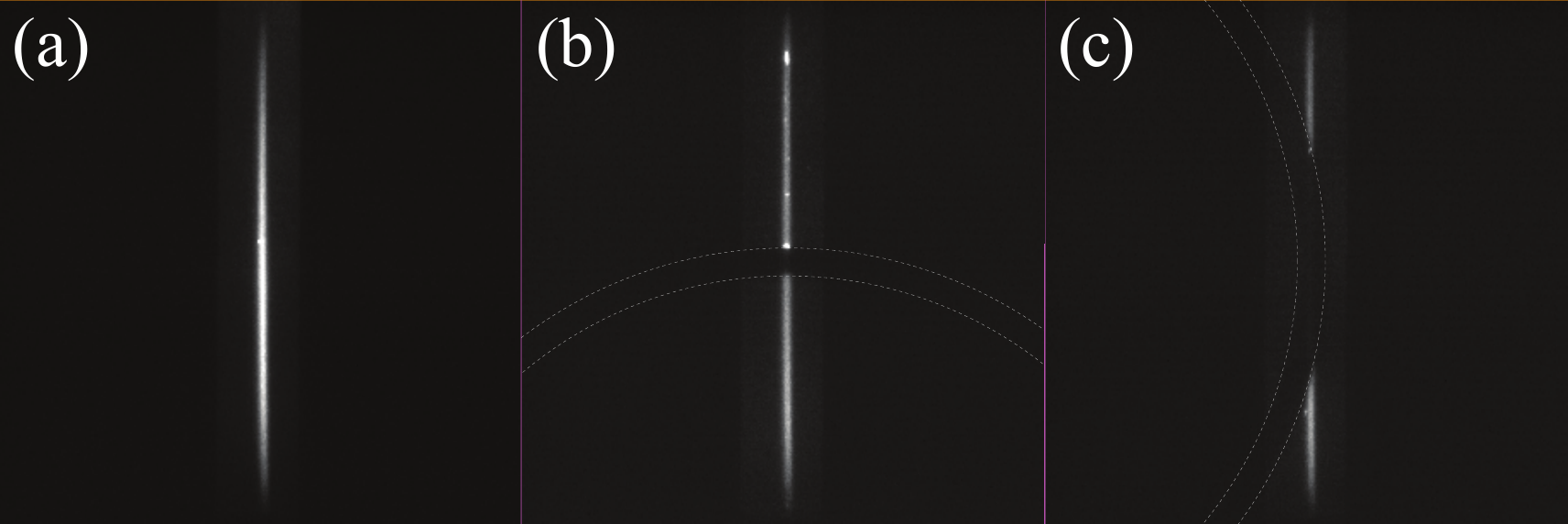}
	\caption{Images of a laser line (a) without a VBG, (b) $\left( \phi,\psi \right)\approx\left(0\degree,6\degree\right)$ and (c) $\left( \phi,\psi \right)\approx\left(6\degree,0\degree\right)$. A collimation lens with a focal length $f = 100$ mm is used. The white dotted lines are added to indicate the rejection ring for better visibility. Note that the image size is 13 mm $\times$ 13 mm and rotated due to the rotated spectrometer and camera assembly.}
	\label{fig:0D1Dexp}
\end{figure}

Figure~\ref{fig:0D1Dexp} demonstrates laserline rejection; (a) without a VBG, (b) with a VBG $\left( \phi,\psi \right)\approx\left(0\degree,6\degree\right)$ and (c) $\left( \phi,\psi \right)\approx\left(6\degree,0\degree\right)$. Note that both collimation lens, L1,  and focusing lens, L2, have a focal length $f = 100$ mm. The white dotted curves are added to indicate the rejection ring for better visibility. Figure~\ref{fig:0D1Dexp}(b) and (c) correspond to the simulated cases shown in Fig.~\ref{fig:0D1D}. 

\begin{figure}
	\includegraphics[width=0.99\linewidth]{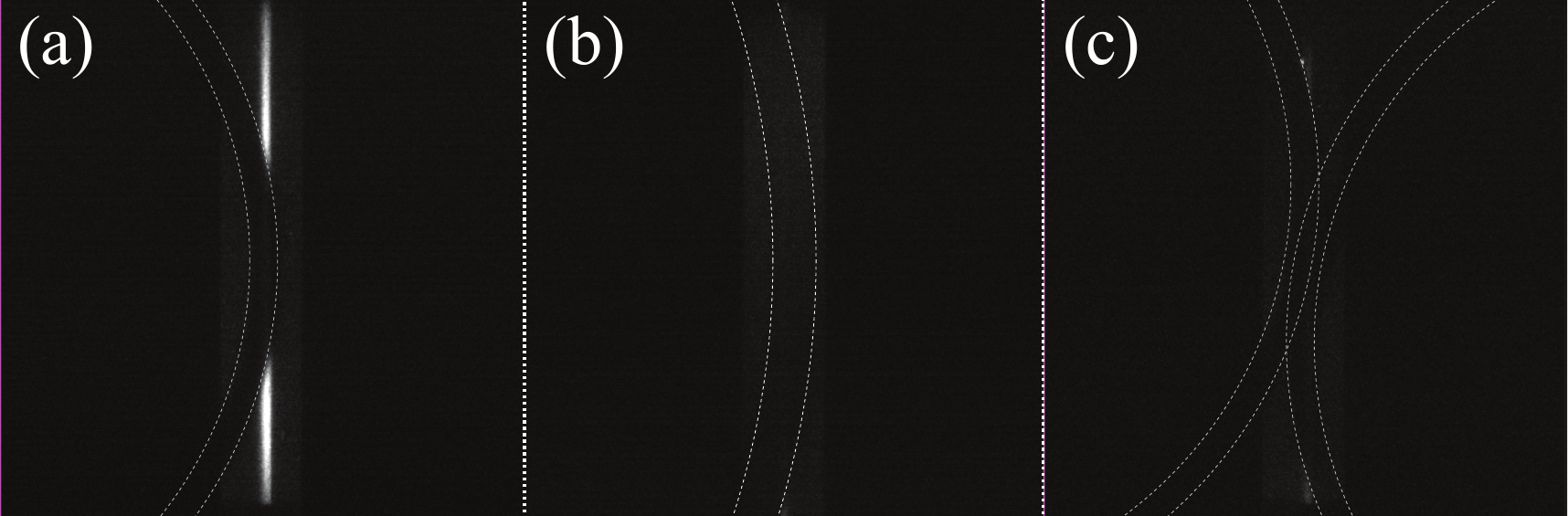}
	\caption{Images of a laser line (a) with one VBG and a collimation lens focal length $f = 100$ mm with $\left( \phi,\psi \right)\approx\left(6\degree,0\degree\right)$, (b) with one VBG and  a collimation lens focal length $f = 200$ mm with $\left( \phi,\psi \right)\approx\left(6\degree,0\degree\right)$. Note that the enlarged rejection ring in (b). In (c), two VBGs are used, where one is angle tuned at $\left( \phi,\psi \right)\approx\left(6\degree,1\degree\right)$ another with $\left( \phi,\psi \right)\approx\left(-6\degree,-1\degree\right)$. The white dotted lines are added to indicate the rejection ring for better visibility.} 
	\label{fig:extenexp}
\end{figure}

Figure~\ref{fig:extenexp} demonstrates laser line rejection; (a) with one VBG and a collimation lens focal length $f = 100$ mm with $\left( \phi,\psi \right)\approx\left(6\degree,0\degree\right)$, (b) with one VBG and  a collimation lens focal length $f = 200$ mm with $\left( \phi,\psi \right)\approx\left(6\degree,0\degree\right)$ and (c) with two VBGs, where one is angle tuned at $\left( \phi,\psi \right)\approx\left(6\degree,1\degree\right)$ another with $\left( \phi,\psi \right)\approx\left(-6\degree,-1\degree\right)$.  The cases of (b) and (c) correspond to the simulated cases in Fig.~\ref{fig:exten}, demonstrating the further extended spatial laser line rejection. 

\section{1D light scattering diagnostics with VBG filters}
Here, we demonstrate 1D spatially-resolved light scattering measurement with VBG filters; rotational Raman scattering and Thomson scattering. A custom vacuum cell and electrode setup is used to control neutral background pressure and discharge voltage conditions. Note that rotational Raman scattering on nitrogen is additionally used for absolute number density calibration of the Thomson scattered light. The general theory and data extraction methods are explained in Appendix.~\ref{sec:scattering}.

Two similar collection systems and test-beds were assembled at Georgia Tech (GT) and Texas A\&M University (TAMU). For both RRS and TS diagnostics, a 1:1 image relay is done with two $f=200$ mm lenses, and then, the 1:2 relay into a spectrometer with a combination of a $f=50$ mm collimation lens and a $f=100$ mm focusing lens. At GT, a single VBG filter is placed between the collimation and focusing lens, while two VBG filters were placed for the experiment at TAMU. For a non-extended 1D measurement, two VBG filters are aligned so that two rejection rings overlap each other for doubled attenuation of the laser line. For an extended 1D measurement, two filters are aligned in a way to reject an extended 1D length.
At GT, a setup composed of a spectrometer (IsoPlane-320A, Princeton Instruments) and an iCCD camera (PI-MAX4:1024i, Princeton Instruments) is used, and at TAMU, the setup described in Sec.~\ref{sec:ringdemo} is used. Both systems employed Thorlabs achromatic lenses for collection, transmission, collimation and focusing on to  the spectrometers. Estimated total transmission from the used collecting optics are 76\% based on the manufactures' specification sheets. The GT experiment uses a Quantel Evergreen 532 nm Nd:YAG laser operating at 10 Hz with 135 mJ of laser energy per pulse as the interrogation beam. The TAMU experiment uses a Ekspla 532 nm Nd:YAG laser operating at 10 Hz with 175 mJ of laser energy per pulse as the interrogation beam. At GT, the nanosecond (ns) pulsed plasma discharge was driven by a ns pulser (NSP-120-20F, Eagle Harbor Technologies). At TAMU, the DC plasma discharge system was driven by a steady 10-kV DC supply (EJ10P60, XP Power). 

\begin{figure}
	\includegraphics[width=0.99\linewidth]{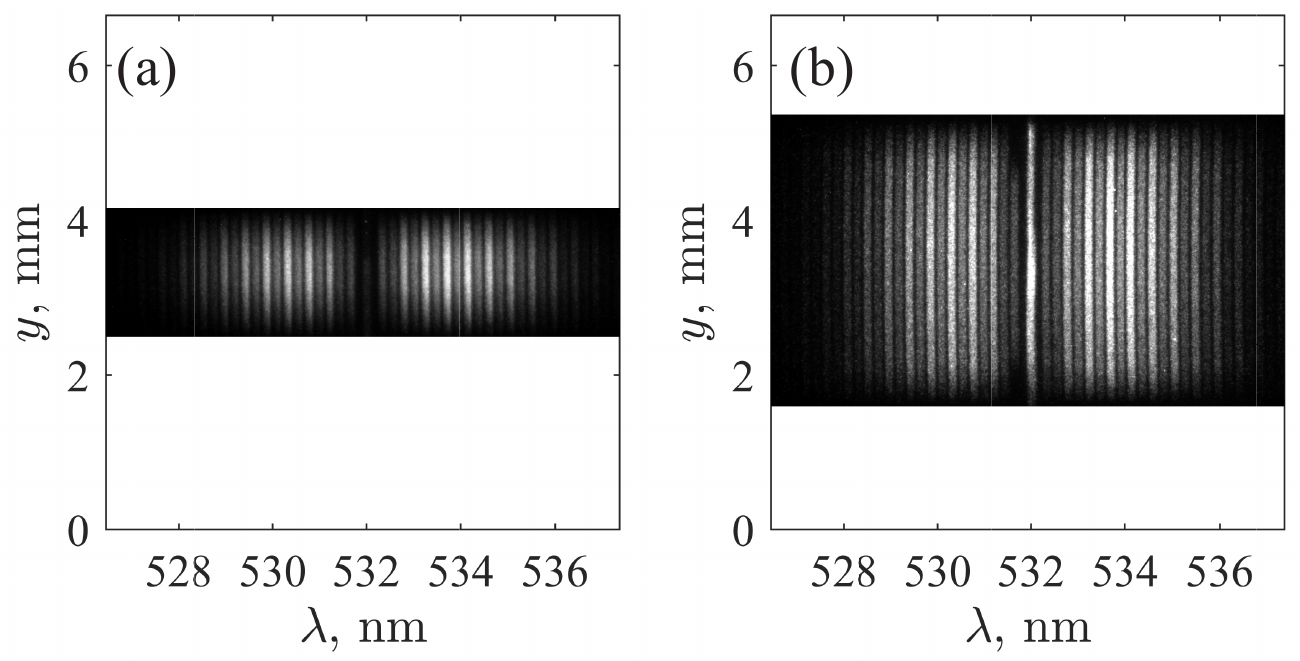}
	\caption{Rotational Raman scattering (RRS) of nitrogen at 150 Torr and room temperature, (a) with one VBG and (b) with two VBGs. The white region is where the view is limited by an extra slit.}
	\label{fig:RRS}
\end{figure}
Figure~\ref{fig:RRS} shows rotational Raman scattering of nitrogen at 150 Torr and room temperature, (a) with one VBG and (b) with two VBGs taken with the system at TAMU, where the laser energy per pulse was 175 mJ and the camera setup was $\times1000$ gain and $200$ accumulation. The domain corresponds to the full camera pixels ($1024\times1024$) and the white region is where the view is limited by an extra slit to physically block the laser line not covered by the VBGs. It is notable that, with an additional VBG, the 1D measurable length is easily extended approximately two times longer than the case with one VBG. 

\begin{figure}
	\includegraphics[width=0.99\linewidth]{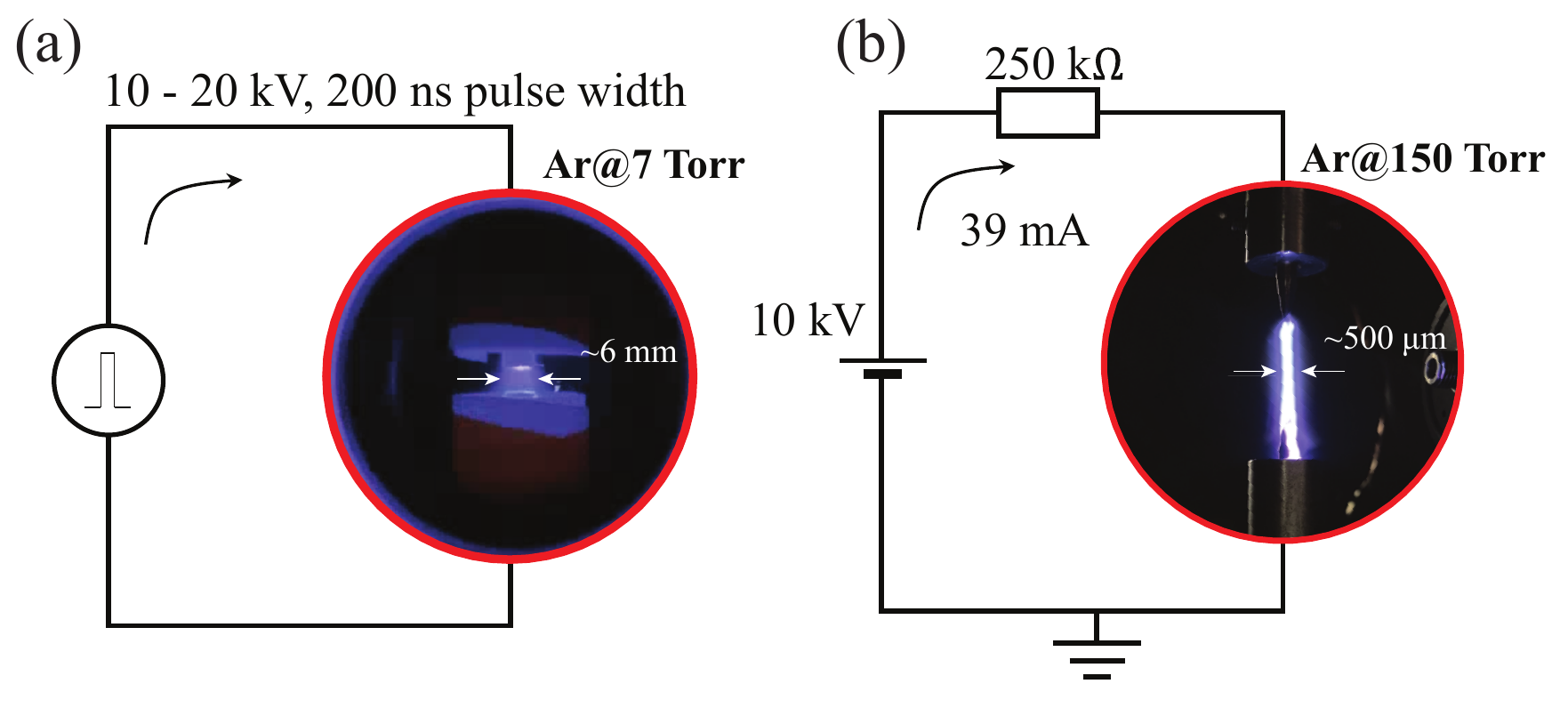}
	\caption{Setups for plasma sources; (a) a nanosecond pulsed discharge setup and (b) a DC discharge setup.}
	\label{fig:discharge}
\end{figure}

The same systems were used for the Thomson scattering detection. Two discharge sources are used; a ns pulsed discharge with argon at 7 Torr and a DC discharge with argon gas at 150 Torr, see Fig.~\ref{fig:discharge}. For TS detection, the used camera setup is $\times100$ gain and $3000$ accumulation for the ns pulsed discharge case, and $\times1000$ gain and $1000$ accumulation for the steady DC discharge case. 

\begin{figure}
	\includegraphics[width=0.99\linewidth]{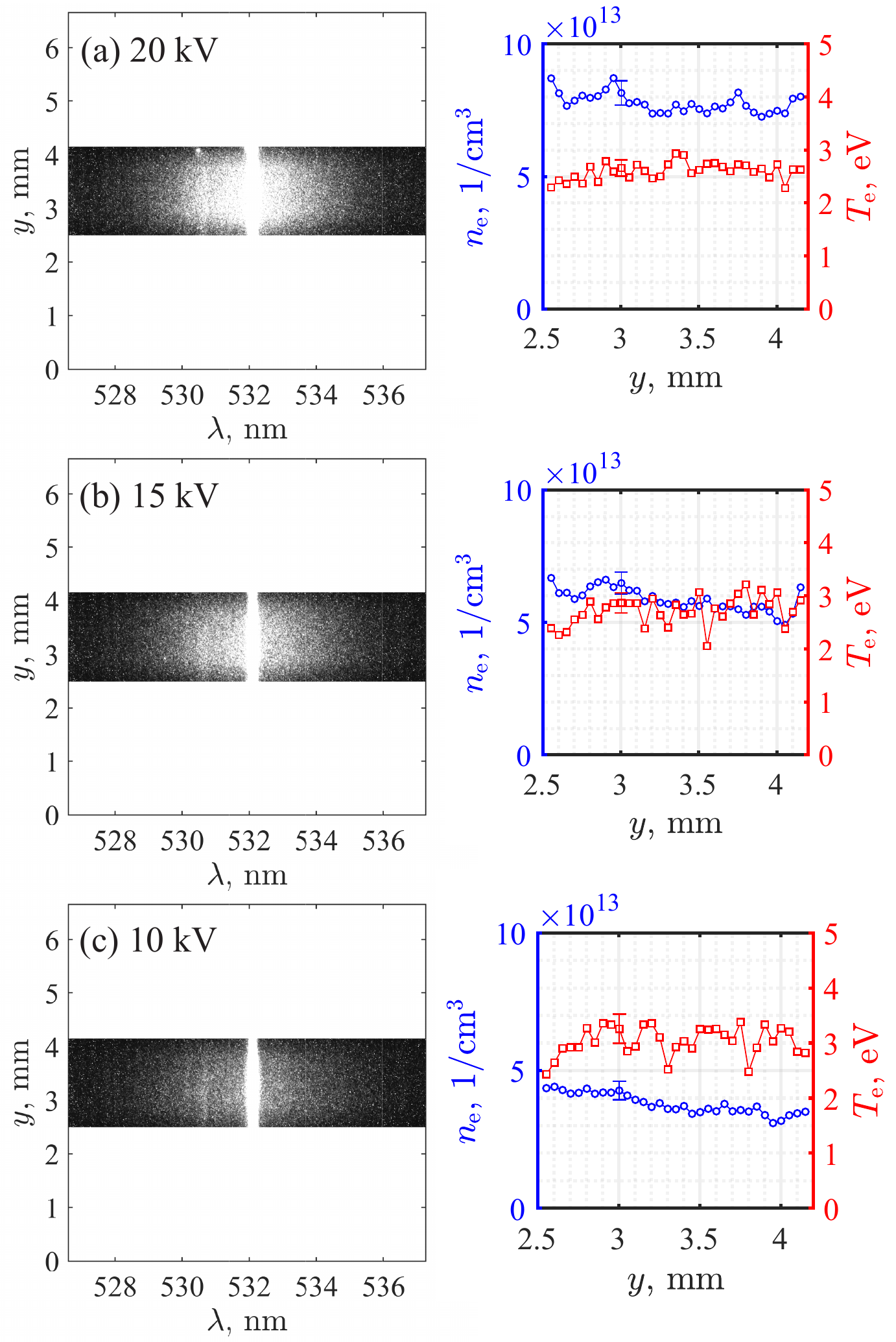}
	\caption{Thomson scattering spectra obtained from the ns pulsed discharge. Total three sample spectra at three different pulse voltages, (a) 20~kV, (b) 15~kV, and (c) 10~kV are shown, and extracted spatial distributions of electron density and temperature are shown on the right hand side where the typical 95\% confidence bounds for the TS spectra fitting are shown with the errorbar.}
	\label{fig:NS}
\end{figure}

Figure~\ref{fig:NS} shows the TS spectra from the ns pulsed discharge. Total three sample spectra at three different pulse voltages, (a) 20~kV, (b) 15~kV, and (c) 10~kV, are shown. The corresponding peak currents at the moment of the scattering collection was 32.1 A, 24.2 A, and 17.8 A, respectively. The measurement region is inside of the plasma column, the size of which is $\sim~6$ mm. Note the darker region at the top and bottom of the spectra comes from various optical aspects such as the vignetting, varying spatial intensity of focused laser, etc. During the density calibration with the nitrogen RRS, spatial distribution of the calibration coefficient is obtained, which prevents a possible artificial underestimation of plasma properties at the boundary that could come from a single calibration coefficient. A local TS spectrum is evaluated to obtain a single point of the distribution curve at every $50~\upmu$m. Note that the spatial resolution in $y$ direction is $6.5~\upmu$m/pixel (1:2 imaging, and the camera resolution of $13~\upmu$m/pixel) without any pixel binning. Eventually, the TS measurement confirms a good spatial uniformity of electron density and temperature, having the coefficient of variation $<0.1$ for the three conditions over the measured length within the ns discharge column. 

\begin{figure}
	\includegraphics[width=0.99\linewidth]{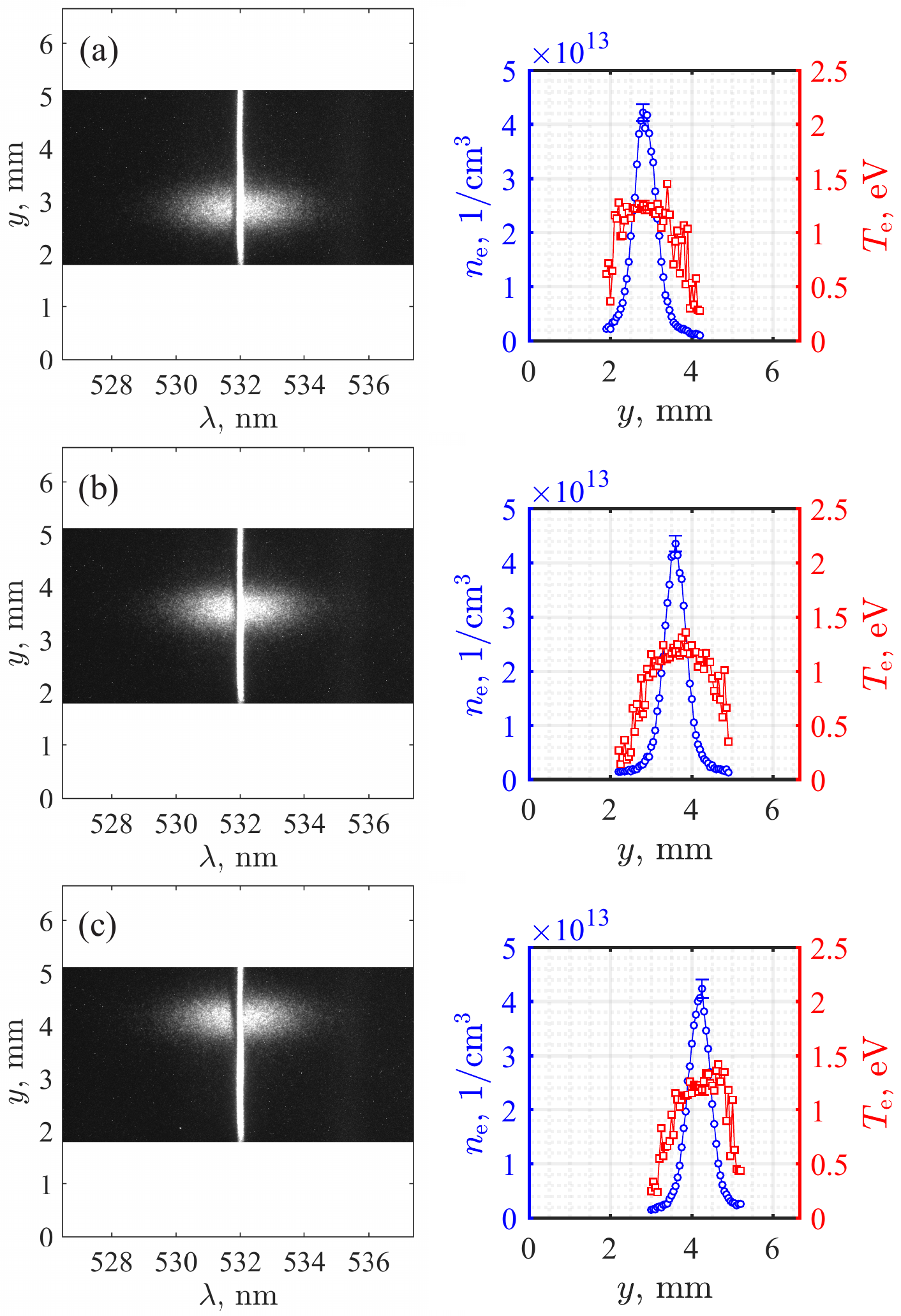}
	\caption{Thomson scattering spectra obtained from the DC discharge. (a-c) shows three different plasma locations within the resolving 1D region. Extracted spatial distributions of electron density and temperature are shown on the right hand side where the typical 95\% confidence bounds for the TS spectra fitting are shown with the errorbar.}
	\label{fig:DC}
\end{figure}

Figure~\ref{fig:DC} shows the TS spectra from the steady DC discharge. To demonstrate the extended 1D TS measurement, the plasma column is spatially shifted while keeping the optics/optical axis fixed. Figure~\ref{fig:DC} (a-c) shows three different plasma locations. Extracted spatial distributions of electron density and temperature are shown on the right-hand side. Again, spatial distribution of the calibration coefficient is obtained during the density calibration with the nitrogen RRS. An example of RRS used for the calibration is that shown in Fig.~\ref{fig:RRS}(b). Similar to the ns discharge experiment, a local TS spectrum is evaluated to obtain a single point of the distribution curve at every $50~\upmu$m. Note that the system has the same spatial resolution of $6.5~\upmu$m explained for the ns discharge. It is noteworthy that the TS is successfully captured at the extended 1D length realized by the two VBG filters. The spatial distribution of electron temperature and density confirms the identical plasma properties detected at different shifted locations. Additionally, thanks to the high resolution of the system, the sharp gradient over a few hundreds $\upmu$m of the plasma column is successfully captured. 

\section{Conclusion}	
We have elucidated the methodology for using a single or multiple VBG notch filters for high spatial resolution of plasma properties perpendicular to the incident interrogation beam.  The systems feature compactness, simple implementation, high throughput, and flexibility to accommodate various experimental conditions. Evaluation the geometrical parameters of a rejection ring by a VBG based on first principles of the VBG acting as a spectral/angular filter is explained. Both simulation and experiment showed how the rejection ring position can be manipulated with a pitch/yaw tuning. It is shown that, with proper angle tuning of the VBG, a system that has traditionally been used for point measurements averaged over the small detection volume necessitated by filter alignment can be easily extended to a 1D spatially-resolved measurement with the exact same setup by simply aligning the filter slightly differently. As the filter is angle sensitive, it could be less effective to reject a stray light which is not coming at a desirable angle to the filter plane. However, an additional optical technique, such as an image relay with a slit, can be implemented.

Several mathematical expressions are derived and presented that allow for estimation of how much spatial rejection and optical attenuation of the laser line can be attainable for a given probe laser beam waist and optical collection system. For the shown example cases with a typical focused probe beam of the beam waist (or the spectrometer slit opening) $\sim 150$~$\upmu$m, the rejection ring can provide several mm length for a 1D measurement.

Methods to further extend spatial measurement region are proposed and demonstrated; using a collimation lens with a different focal length or using multiple VBG filters. The method using multiple filters can minimize the signal loss, without the trade-off of the solid angle. Note that this proposed use of multiple VBG filters differs from the commonly known application to improve the total attenuation by extending the possible detection volume in the beam propagation axis direction.

Extended spatial measurement is greatly advantageous as it provides further comprehensive spatial information of the test articles, such as spatial uniformity/variation of physical properties. The two different plasma sources are used, and the system successfully observed the spatial uniformity within a $\sim6$~mm column, or the sharp gradient over a few hundreds $\upmu$m with a $6.5~\upmu$m spatial resolution. It should be noted that this resolution can be easily tuned by imaging optics and pixel binning. The proposed system can be helpful to characterize plasmas which have small plasma size of a few mm or steep gradients\cite{Sasaki2015ElectronMPa,Hubner2017ThomsonJets,Babaeva2021UniversalMedia,Johnson2020Spatio-temporalSubstrates,Alberti2020CollinearDeposition,Darny2021PeriodicJet,Viegas2022PhysicsExperiments,Singh2022InsightEnergy}. Note that the capability to extend 1D measurement is also highly useful to diagnose larger scale test articles as well.

\appendix
\section{\label{sec:scattering}Scattering of Electromagnetic Radiation and data analysis}
Scattering of electromagnetic radiation is the process of absorption and re-radiation of incident electromagnetic radiation by molecules. This process can be elastic, leading to a re-radiation at the same wavelength/frequency of the incident radiation in the frame of reference of the absorbing molecule. This process can also be inelastic, leading to re-radiation that is at a different wavelength/frequency than that of the incident radiation in the frame of reference of the scatterer. This causes an energy transition in one of the energy modes of the molecules undergoing the scattering process. 

There are many types of electromagnetic scattering. The relevant ones in this work are Thomson scattering and rotational Raman scattering. Thomson scattering is the elastic scattering of incident radiation from free unbound charges. Both ions and electrons Thomson scatter, but the ion scattering cross section is several orders of magnitude smaller than that of the electron scattering cross section due to their relative masses, and as such, very difficult to measure. In this work, we are concerned with electron Thomson scattering. 

Raman scattering is the inelastic scattering of polyatomic molecules as a result of a net energy exchange between the incident radiation and the internal energy modes of the polyatomic molecule. Raman scattering can be purely rotational, purely vibrational, purely electronic, or a combination of any of the aforementioned. In this work we are concerned with purely rotational Raman scattering, as this occurs in the frequency range closest to the incident frequency.

The quantification of the scattered power or photon count given an incident laser beam has a simple functional form, given by 
\begin{equation}
P_\mathrm{s}(\lambda) = C_{1} [(E_\mathrm{i} L_\mathrm{det} \Delta \Omega ) n \frac{\partial \sigma}{\partial \Omega}]  S(\lambda),
\label{eq:scat1}
\end{equation}
where the constant $C$ represents an efficiency constant, $E_\mathrm{i}, L_\mathrm{det},\Delta \Omega, n$ being the incident laser energy, detection volume length and detection solid angle, and scatterer number density, respectively. $\frac{\partial \sigma}{\partial \Omega}$ is the scattering cross section. The items in the bracket represent the total scattered power intensity. $S(\lambda)$ represents the spectral shape function, which results from the finite nature of the detection optics, light dispersion equipment (in our case, spectrometer), and detector. This function serves to redistribute the total scattered power and follows all of the properties of a probability distribution function in the system of units chosen. Below, we present the forms of the Raman and Thomson scattering equations. See Ref.~\cite{Sande2002LaserRejection,Vincent2018AStudies,Carbone2015ThomsonChallenges} for more details.

So long as the collection conditions remain fixed, this equation can be simplified to
\begin{equation}
P_\mathrm{s}(\lambda) = Cn \frac{\partial \sigma}{\partial \Omega} S(\lambda).
\label{eq:scat2}
\end{equation}

\subsection{\label{sec:ramanscattering}Raman Scattering}
In the case of rotational Raman scattering (RRS), the equation can be specified to
\begin{equation}
P^\mathrm{R}_\mathrm{s}(\lambda) = C \sum_{J'=J\pm2}n_{J} \frac{\partial \sigma^\mathrm{R}_{J \rightarrow J'}}{\partial \Omega} S^\mathrm{R}(\lambda-\lambda_{J \rightarrow J'}), 
\label{eq:scat3}
\end{equation}
where $n_J$ is the density for each state $J$ given as,
\begin{equation}
{n_J} = \frac{n_\mathrm{g}}{Q_\mathrm{R}}g_J(2J+1)\exp[-\frac{\epsilon_J(J)}{k_\mathrm{B}T_\textbf{g}}].
\end{equation}
Note that $n_\mathrm{g}$ is the neutral gas number density, $Q_\mathrm{R} = \frac{(2I+1)^2k_\mathrm{B} T_\mathrm{g}}{2Bhc}$ the partition sum, $\epsilon_J(J) = hcBJ(J+1)$ the state energy, and $g_J$ the statistical weighting factor. For nitrogen, $g_J$ is either 6 if $J$ is even or 3 if $J$ is odd. $I$ is unity, $B$ is 198.973 $\mathrm{m}^{-1}$. $S^\mathrm{R}(\lambda-\lambda_{J \rightarrow J'})$ represents the shape profile taking into account various spectrum broadening sources. In RRS, it is generally assumed that the only source of broadening is the instrument function (IF), and is modeled as a Gaussian function with a characteristic full width at half max (FWHM), which is related to the Gaussian standard deviation $\sigma$ by $\mathrm{FWHM}=2\sqrt{2\ln2}\sigma$. We can quantify this as 
\begin{equation}
S^\mathrm{R}(\lambda,\sigma_\mathrm{IF}) = \frac{1}{\sqrt{2\pi}\sigma_\mathrm{IF}}\exp(-\frac{1}{2}\frac{(\lambda - \lambda_{J \rightarrow J'})^2}{\sigma^2_\mathrm{IF}}),
\end{equation}
which is the outcome of the convolution between the Dirac delta function $\delta(\lambda - \lambda_{J \rightarrow J'})$ and the Gaussian shape function $S(\lambda)$ and is mathematical representation of the redistribution of the rotational Raman intensity lines due to the finite detection system. 

The differential Raman cross-section $\frac{\partial \sigma^{\mathrm{R}}}{\partial \Omega}$ is further broken up into its stokes and anti-stokes bands, which correspond to transitions to wavelengths above and below the incident wavelength. 

Each of these takes the following functional form
\begin{equation}
\frac{\partial \sigma^{\mathrm{R}}_{J \rightarrow J'}}{\partial \Omega} = [\frac{3}{4}]\frac{\partial \sigma^{\mathrm{R}}_{\bot }}{\partial \Omega},
\end{equation}
where the factor $3/4$ is coming from the depolarization of RRS\cite{Penney1974AbsoluteCO_2} and the perpendicular cross-section is given as,
\begin{equation}
\frac{\sigma^\mathrm{R}_\bot}{\partial \Omega} = [\frac{64\pi^4}{45}\frac{\gamma^2}{\epsilon_\mathrm{0} ^2}][\frac{b_{J \to J'}(J)}{\lambda^4_{J \to J'}(J)}].
\end{equation}
Here, $\gamma$ is the anisotropy of the molecular polarisability tensor, and the Placzek-Teller coefficient $b_{J \to J\pm2}$ and the scattered wavelength $\lambda_{J \to J\pm2}$ are given as
\begin{equation}
b_{J \to J\pm2}(J) = \frac{3(2J+1\pm 1)(2J+1 \pm 3)}{4(2J+1)(2J+1 \pm 2)},
\end{equation}
and
\begin{equation}
\lambda_{J \to J\pm2}(J) = \lambda_\mathrm{i} \pm  \lambda_\mathrm{i}^2 \frac{B}{hc}(4J+2\pm4),
\end{equation}
respectively. The '$\pm$' sign corresponds to the stokes and anti-stokes portions of the spectrum respectively. 

Given a Raman spectrum collected at a given number of accumulations, detector gate, laser energy, at a known pressure and temperature, a fitting procedure can be used to determine $C$ and $\sigma_\mathrm{IF}$ as a function of spatial location. Collecting the Thomson spectra at the same collection conditions allows for $C$ to be used as the calibration constant for the absolute number density. 

\subsection{\label{sec:thomsonscattering}Thomson Scattering}
In the case of Thomson scattering (TS)
\begin{equation}
P^\mathrm{T}_\mathrm{s}(\lambda) = C [n_\mathrm{e} \frac{\partial \sigma^\mathrm{T}}{\partial \Omega}]  S^\mathrm{T}(\lambda),
\label{eq:scat4}
\end{equation}
with the cross section being equal to the classical electron radius as
\begin{equation}
\frac{\partial \sigma^\mathrm{T}}{\partial \Omega} = r_\mathrm{e}^2\equiv \frac{e^2}{4\pi\epsilon_\mathrm{0}m_\mathrm{e}c^2}.
\end{equation}
Assuming a Maxwellian velocity distribution and neglecting the instrument broadening on the Thomson spectrum as is typically done, the redistribution operation is a simple multiplication by the following:
\begin{equation}
S^\mathrm{T}(\lambda)=\frac{\partial \omega}{\partial \lambda} S^\mathrm{T}(\omega),
\end{equation}
where
\begin{equation}
S^\mathrm{T}(\omega) = \sqrt{\frac{m_\mathrm{e}}{2\pi k_\mathrm{B} T_\mathrm{e} k^2}} \exp(\frac{1}{2}\frac{(\omega_\mathrm{i}-\omega_\mathrm{s}-kv_\mathrm{d})^2}{2k_\mathrm{B}T_\mathrm{e}k^2}/m_\mathrm{e}),
\end{equation}
with
\begin{equation}
\omega = \frac{2\pi c}{n}\frac{1}{\lambda},
\end{equation}
\begin{equation}
\frac{\partial \omega}{\partial \lambda} = \frac{2\pi c}{n}\frac{1}{\lambda ^2},
\end{equation}
\begin{equation}
k^2 = k_\mathrm{s}^2+k_\mathrm{i}^2-2k_\mathrm{i}k_\mathrm{s},
\end{equation}
\begin{equation}
k=\frac{2\pi}{\lambda}.
\end{equation}
The $\partial \omega / \partial \lambda$ is necessary to preserve the probability density function integral being unity when being cast in the wavelength domain. With $C$ from the Raman calibration, this can easily be fit to determine $n_\mathrm{e}$ and $T_\mathrm{e}$ for a given Thomson spectrum collected with the same collection conditions.

\section*{References}
\bibliography{references}
\bibliographystyle{iopart-num}
\end{document}